\documentstyle[graphicx,prb,multicol,aps]{revtex}
\newcommand{\lsim}{{ _< \atop ^\sim}}
\newcommand{\gsim}{{ _> \atop ^\sim}}
\newcommand{\be}{\begin{eqnarray}}
\newcommand{\ee}{\end{eqnarray}}

\begin{document}

\title{4f--spin dynamics in $\rm\bf La_{2-x-y}Sr_xNd_yCuO_4$}
\author{$\rm ^1 M.$ Roepke, $\rm ^1 E.$ Holland--Moritz, $\rm ^1 B.$ B\"uchner,
$\rm ^1 H.$ Berg, $\rm ^2 R.E.$ Lechner, $\rm ^2 S.$ Longeville,
$\rm^2J.$ Fitter, $\rm ^3 R.$ Kahn,\linebreak
$\rm ^3 G.$ Coddens and $\rm ^4 M.$ Ferrand}
\address{$^1 II.$ Physikalisches Institut, Universit\"at zu K\"oln, Z\"ulpicher Str. 77,
D--50937 K\"oln}
\address{$^2 H$ahn--Meitner--Institut, Glienicker Str. 100, D--14109 Berlin}
\address{$^3 Laboratoire$ Leon Brillouin, CEA--Saclay, 91191 Gif--sur--Yvette Cedex,
France}
\address{$^4 Institut$ Laue--Langevin, Ave. des Martyrs, BP 156, 38042 Grenoble
Cedex 9, France}
%
%\date{\today}
%
\maketitle
\begin{abstract}

We have performed inelastic magnetic neutron scattering
experiments on $\rm La_{2-x-y}Sr_xNd_yCuO_4$ ($\rm 0 \leq x \leq 0.2$ and $\rm 0.1
\leq y \leq 0.6$) in order to study the Nd 4f--spin dynamics at low energies
($\rm\hbar\omega~ \lsim ~ 1~meV$). In all
samples we find at high temperatures a quasielastic line
(Lorentzian) with a line width which decreases on lowering the
temperature. The temperature dependence of the quasielastic line width $\rm
\Gamma/2(T)$ can be explained with an Orbach--process, i.e. a relaxation via
the coupling between crystal field excitations and phonons. At low temperatures
the Nd--4f magnetic response $\rm S({\bf Q},\,\omega)$
correlates with the electronic properties of the $\rm CuO_2$--layers. In the
insulator $\rm La_{2-y}Nd_yCuO_4$ ($\rm y = 0.1,\, 0.3$) the quasielastic
line vanishes below  80~K and an inelastic excitation occurs. This directly
indicates the splitting of the $\rm Nd^{3+}$ ground state Kramers doublet due
to the static antiferromagnetic order of the Cu moments. In
$\rm La_{1.7-x}Sr_xNd_{0.3}CuO_4$ with x = 0.12, 0.15 and $\rm
La_{1.4-x}Sr_xNd_{0.6}CuO_4$ with x = 0.1, 0.12, 0.15, 0.18 superconductivity
is strongly suppressed. In
these compounds we observe a temperature independent broad quasielastic line
of Gaussian shape below
$\rm T \approx 30~K$. This suggests a distribution of various internal fields
on different Nd sites and is interpreted in the frame of the stripe model.
In $\rm La_{1.8-y}Sr_{0.2}Nd_yCuO_4$ ($\rm y =
0.3,\, 0.6$) such a quasielastic broadening is not observed even at lowest
temperature.

\end{abstract}

\vspace*{-11.5cm}\noindent {\sc Physical Review B} \hfill to be published
1 October 1999
\vspace*{11cm}
\begin{multicols}{2}
%
%\narrowtext
%
%

\section{Introduction}

In the high $\rm T_c$ superconductors a close interplay between
superconductivity and magnetism exists. The parent compounds of the
cuprates are antiferromagnetic insulators. Doping with holes or electrons
destroys the long range order but antiferromagnetic correlations persist
even in the superconducting region. {\em Inelastic} incommensurate magnetic
peaks in superconducting $\rm La_{2-x}Sr_xCuO_4$ (LSCO) show that
superconductivity and magnetic correlations coexist\cite{Incommens}. This
phenomenon has regained attraction since in 1995 {\em elastic} peaks at the
same incommensurate positions were found in $\rm
La_{1.48}Sr_{0.12}Nd_{0.4}CuO_4$. These elastic peaks are interpreted in
terms of hole--rich and spin--rich domains in the $\rm CuO_2$--layers, i.e.
the well known stripe picture\cite{Tranquada1}. In LSCO (T--phase) doping
with Nd induces a further low temperature phase transition for $\rm Nd\geq
0.18$. For x = 0 there is a transition from the low temperature
orthorhombic LTO to the less orthorhombic Pccn phase\cite{Crawford} whereas
for $\rm x~\gsim~0.1$ the transition is to the tetragonal LTT phase. In the
latter superconductivity is strongly suppressed for certain Sr
concentrations\cite{Buechner}. In the LTT phase the tilt of the $\rm
CuO_6$--octahedra can serve as pinning potential for the dynamical
stripe correlations. Hence, the inelastic peaks become elastic indicating a
formation of {\em static} anti phase antiferromagnetic domains in the $\rm
CuO_2$--planes which are separated by quasi one--dimensional stripes
containing the doped charge carriers. Recently, inelastic incommensurate
peaks have also been observed in superconducting $\rm
YBa_2Cu_3O_{6.6}$\cite{Dai} giving rise to the question whether stripes are
a general feature of cuprate based high $\rm T_c$ superconductors.\par
In addition to the investigation of the Cu subsystem, the dynamics of the Nd
spins in $\rm Nd_{2-x}Ce_xCuO_4$ (NCCO, $\rm T^{'}$--phase) has been examined
extensively by several
groups\cite{Henggeler1,Ivanov,Casalta,Loewenhaupt1}. Henggeler et
al.\cite{Henggeler2}
performed neutron scattering experiments to examine the spin excitation
spectrum at low temperatures. The same group explained the heavy fermion like
large $\rm \gamma$ coefficient in low temperature specific heat
measurements\cite{Maiser} of NCCO by the shift of spectral weight of the Nd
modes to lower energies with increasing number of charge
carriers\cite{Henggeler3}. Specific heat measurements show a
Schottky anomaly in the parent compound $\rm Nd_2CuO_4$\cite{Markert}. This is
explained by the presence of Nd--Cu interactions being responsible for the
splitting of the $\rm Nd^{3+}$ ground state Kramers doublet as e.g. observed in
Raman and neutron scattering
experiments\cite{Jandl1,Dufour,Jandl2,Loewenhaupt2}.
At higher temperatures in neutron scattering experiments on powder samples of
NCCO\cite{Loewenhaupt2} and on a single crystal of $\rm Nd_2CuO_4$\cite{Casalta}
a quasielastic (QE) Lorentzian is observed with a line width that increases almost
linearly with increasing temperature.
At lower temperatures the line shape turns into a Gaussian with an almost
constant line width. Similar features have been presented for $\rm
NdBa_2Cu_3O_{7-\delta}$\cite{Droessler}. In this compound the QE Gaussian
and the Lorentzian coexist.\par
In the present work we report on inelastic magnetic neutron scattering
experiments on $\rm La_{2-x-y}Sr_xNd_yCuO_4$ at various
temperatures. We have investigated the 4f magnetic response for samples with
Sr--concentrations $\rm 0 \leq x \leq 0.2$ and Nd--concentrations $\rm 0.1
\leq y \leq 0.6$ at low
energies (typically $\rm \sim 1~meV$) in order to obtain information about
the Cu magnetism in the $\rm CuO_2$--layers via the Nd--Cu interaction. The
paper is organized as \mbox{follows\,:} the
next chapter describes the experimental technique. The presentation
of our results and their discussion follows in chapter III which is divided
into two parts as justified by our experimental findings. In section
IV we will give a brief summary.

\section{Experimental}

We performed temperature dependent studies on $\rm La_{1.9}Nd_{0.1}CuO_4$,
$\rm La_{1.7-x}Sr_xNd_{0.3}CuO_4$ (x = 0, 0.12, 0.15, 0.2) and
$\rm La_{1.4-x}Sr_xNd_{0.6}CuO_4$ (x = 0.1, 0.12, 0.15, 0.18, 0.2)
using the time--of--flight (TOF)
spectrometers V3 NEAT\cite{Lechner} (HMI Berlin), G6.2 MIBEMOL (LLB
Saclay) and IN5 (ILL Grenoble). All spectrometers are located at cold
neutron beam lines and use chopper systems for
monochromatization which give very sharp and clean resolution functions. The
energy chosen for the incident neutrons ranges from $\rm
E_i=3.15~meV$ ($\rm\equiv 5.1\AA$) to $\rm E_i=1.28~meV$ ($\rm\equiv 8\AA$)
resulting in energy resolutions between $\rm \Delta E\approx50~\mu eV$ and $\rm
\Delta E\approx20~\mu eV$ (HWHM), respectively. Additionally, $\rm
La_{1.45}Sr_{0.15}Nd_{0.4}CuO_4$ was measured at the IN6 (ILL, Grenoble) with
$\rm E_{i}=3.15~meV$ ($\rm \Delta E\approx45~\mu eV$)\cite{Report}. This
spectrometer uses Bragg diffraction on single
crystals to monochromize the neutron beam. In all cases a Vanadium standard for
calibration and an empty can measurement of the Al flat plate for background
correction were carried out. For the experiments we used well characterized
powder samples\cite{Breuer} of typically m = 20~g and a standard cryostat
for cooling. Details of the data analysis are described
elsewhere\cite{Holland}. Since we could not find any
Q--dependence of the magnetic signals $\rm S(Q,\,\omega)$ is averaged over
a broad Q--window for all spectra to obtain a better statistics.

\section{Results and Discussion}

\subsection{4f spin relaxation due to coupling of phonons and crystal field
excitations}

\begin{figure}[h]
\begin{center}
\includegraphics[clip, angle =270,width=8.5cm]{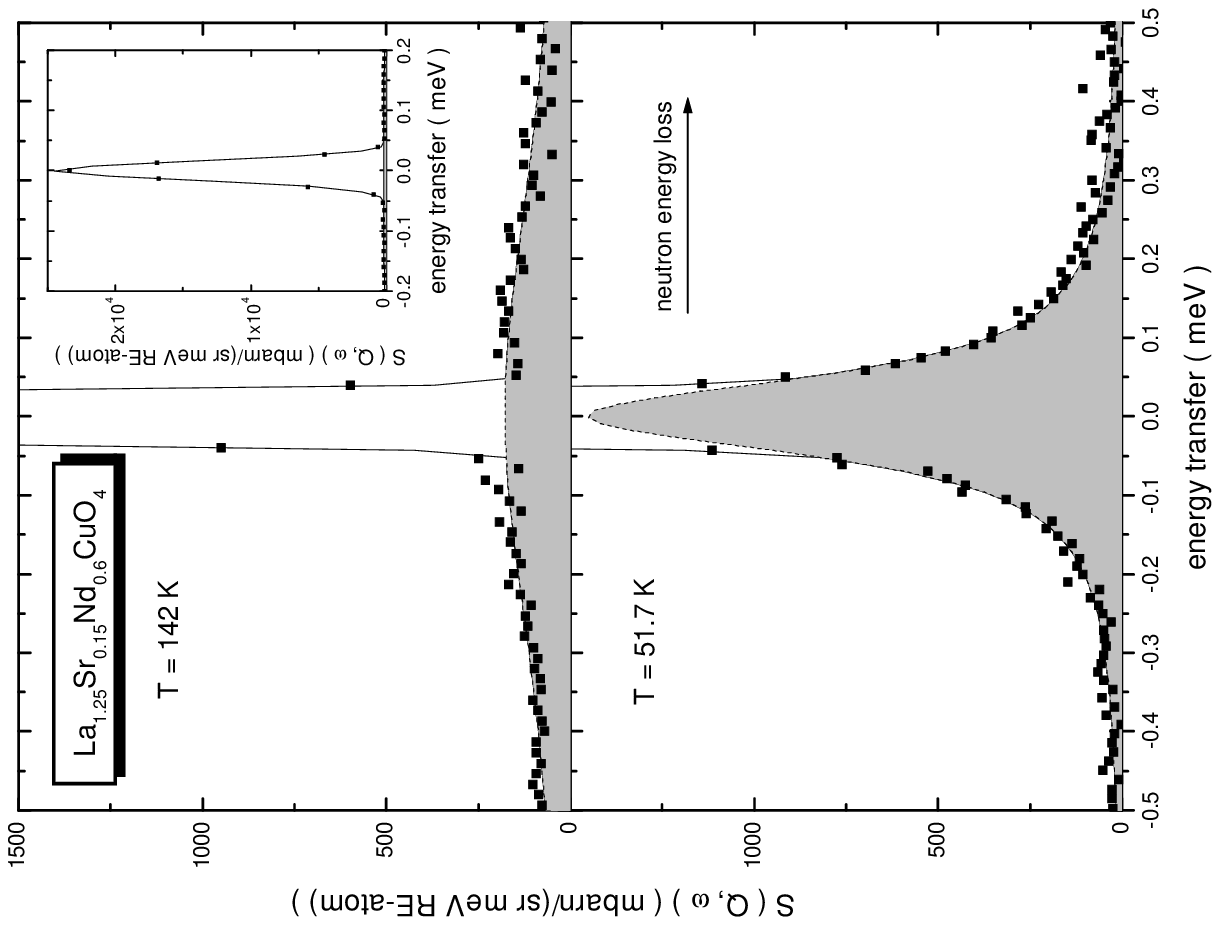}
\end{center}
\vspace*{-.15cm}
\begin{minipage}{8.5cm}
\caption{\it Background\hfill corrected\hfill spectra\hfill of\protect\newline
$La_{1.25}Sr_{0.15}Nd_{0.6}CuO_4$ (IN5). The full line is the best fit
including the nuclear incoherent elastic scattering. The magnetic contribution
(QE Lorentzian convolved with the instrumental resolution function) is
illustrated by the shaded area. The inset shows the overall elastic
scattering.}
\end{minipage}
\label{prb001}
\end{figure}

\begin{figure}
\begin{center}
\includegraphics[clip, angle =270,width=8.5cm]{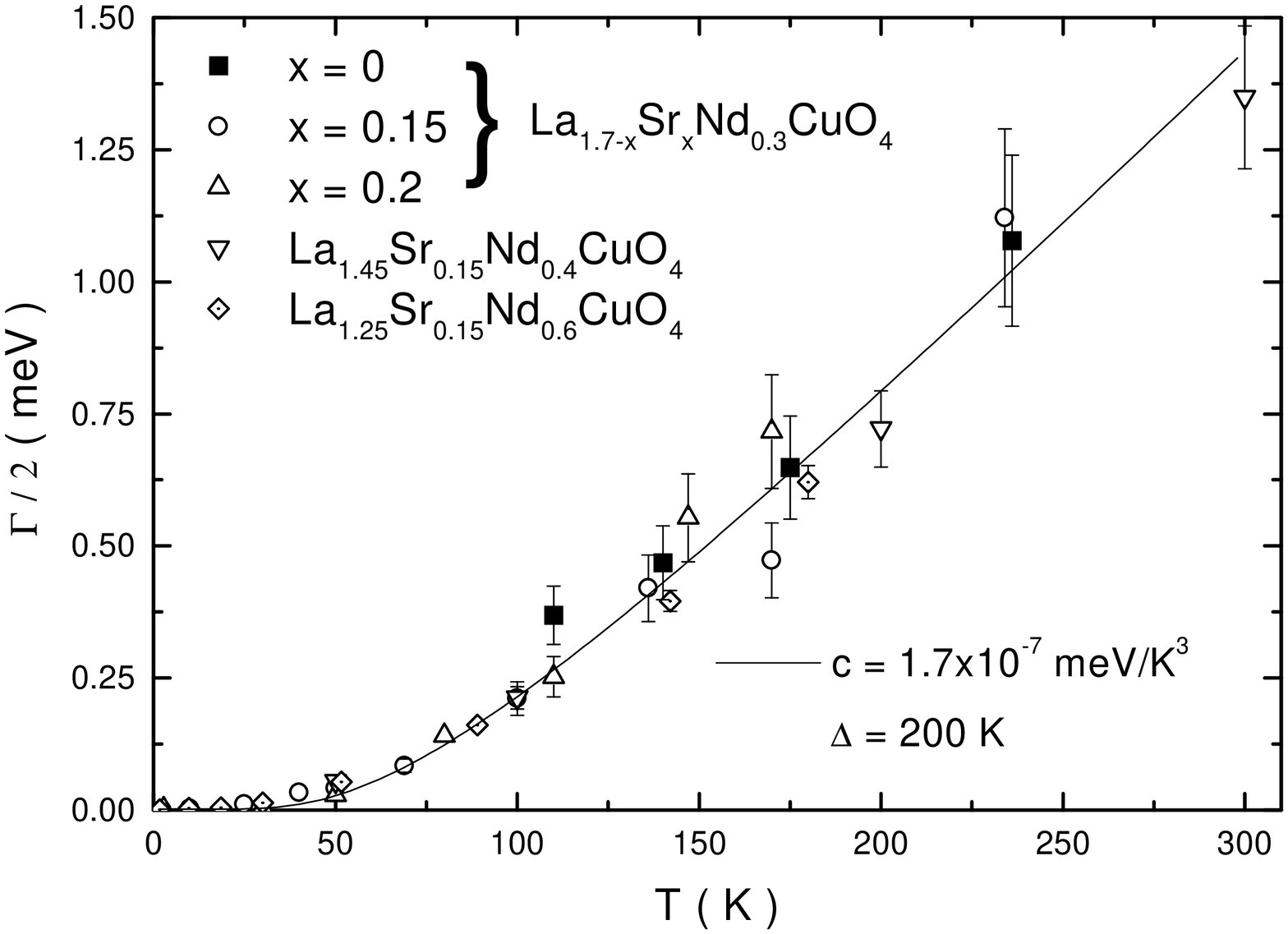}
\end{center}
\vspace*{-.15cm}
\begin{minipage}{8.5cm}
\caption{\it Quasielastic line widths of $La_{1.7-x}Sr_xNd_{0.3}CuO_4$ with x =
0, 0.15, 0.2 (NEAT), $La_{1.45}Sr_{0.15}Nd_{0.4}CuO_4$ (IN6) and $
La_{1.25}Sr_{0.15}Nd_{0.6}CuO_4$ (IN5). The full line is a function according to
equation (1) with a mean value for parameter c (see text).}
\end{minipage}
\label{prb002}
\end{figure}

In all samples a quasielastic line of Lorentzian shape is observed at high
temperatures. This is illustrated in Fig. \ref{prb001} showing spectra of
$\rm La_{1.25}Sr_{0.15}Nd_{0.6}CuO_4$ for two different temperatures. The line
width $\rm \Gamma/2(T)$ of the
Lorentzian decreases with decreasing temperature and shows the {\em same}
temperature dependence in {\em all} compounds within experimental error.
In Fig.
\ref{prb002} the temperature dependence of the QE line width for samples with
different Sr-- and Nd--concentrations is plotted. Above about 100~K the line
width increases almost linearly with increasing temperature. Below
this temperature the slope is drastically reduced although the line width
decreases furthermore on lowering the temperature. The residual line width
for $\rm T \to 0~K$ is below the resolution limit even in the experiments when
an energy resolution of $\rm 20~ \mu eV$ (HWHM) was chosen. Generally, it is
hard to detect the QE Lorentzian for $\rm T\,\lsim \,20~K$. However, when we do not
include a QE Lorentzian in our fitting procedure we observe an increase of the
elastic intensity in this temperature region. Since the coherent and incoherent
elastic scattering intensity should be almost temperature independent for each
sample we can conclude that this additional intensity originates from
magnetic scattering. If the fit for the QE Lorentzian yielded a value smaller
than half of the resolution (HWHM) we fixed the line width at $\rm \Gamma/2=0$
for this temperature.

The QE line width is related to the spin fluctuation frequency via $\rm \Gamma/2
=\hbar / \tau$. Thus the decrease of the line width with decreasing temperature
is a direct evidence for the lowering of the 4f--spin fluctuation frequency.\par
The observation of a magnetic QE line in high $\rm T_c$ superconductors and
related materials has been a subject for discussion for over a
decade\cite{Loewenhaupt2,Droessler,Walter,Allenspach,Allenspach2}.
During this time two attempts to describe the 4f spin relaxation, i.e. the
temperature dependence of the QE line width, were suggested\,:
\begin{itemize}
\item[(i)] via the interaction of 4f spins with conduction electron spins
\item[(ii)] by exchange interaction between 4f spins themselves
\end{itemize}
In the first case the line width should increase linearly with temperature as
$\rm \Gamma/2 \propto \left[N(\epsilon_F) \cdot J_{ex}\right]^2 \cdot T$
(Korringa law) with $\rm N(\epsilon_F)$ the density of electron states at the
Fermi energy
and $\rm J_{ex}$ the exchange integral between 4f-- and conduction electron
spins\cite{Becker}. Such a temperature dependence was
often found in intermetallic systems\cite{Fulde}. In the second
case a power law is
expected\cite{Staub2}.

\begin{figure}
\begin{center}
\includegraphics[clip, angle =0,width=8.5cm]{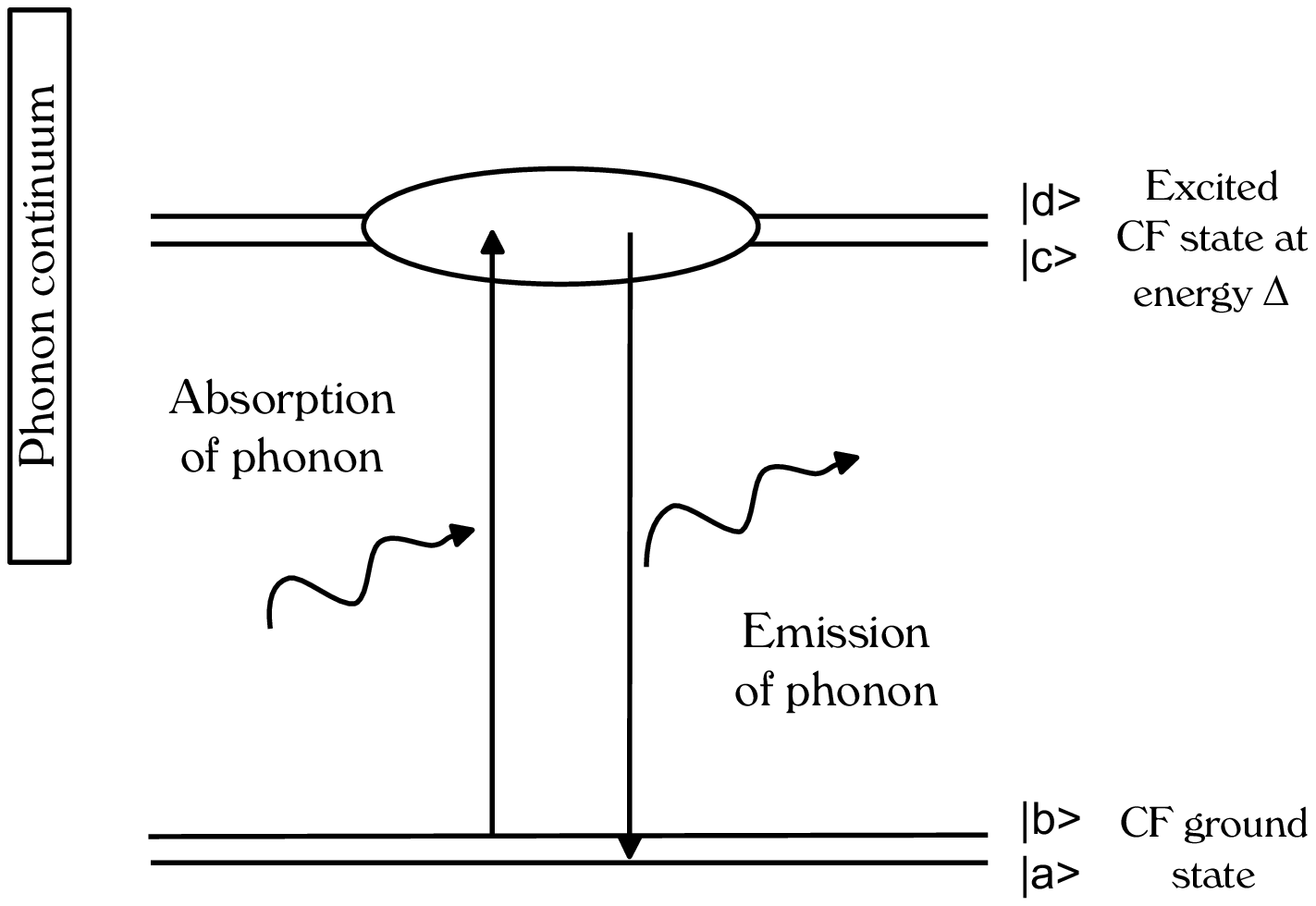}
\end{center}
\vspace*{-.15cm}
\begin{minipage}{8.5cm}
\caption{\it Schematic diagram of the Orbach relaxation process. $|\,a> \,
and \; |\,b>$ denote the two states of the CF ground state,
$|\,c>\;and \; |\,d>$ of an excited state, respectively (for a Kramers ion like
$Nd^{3+}$). The arrows indicate a
possible transition with involvement of a phonon.} \label{orbachproz}
\end{minipage}
\end{figure}

From our data we can exclude both cases, since we
observe the same temperature dependence of $\rm \Gamma/2(T)$ in samples with
($\rm x>0$) and without charge carriers ($\rm x=0$) which rules out scenario (i).
Again, the same temperature dependence of $\rm \Gamma/2(T)$ is obtained
for samples with high Nd content (y = 0.6) and low Nd content (y = 0.1). For a
small Nd concentration the Nd--Nd exchange interaction should be weak, which
rules out scenario (ii).
After a detailed analysis of our data we found that the relaxation of the
4f--spins can be well described with a two--phonon Orbach process\cite{Orbach}
assuming a coupling between crystal field (CF) excitations and phonons. The
Orbach process is sketched in Fig. \ref{orbachproz}. In this relaxation process
a Nd ion which is in state $\rm |\,b>$ can absorb a phonon and hence is excited to
state $\rm |\,c>$ (or $\rm |\,d>$). From this state it can fall down to state
$\rm |\,a>$ by emitting a second phonon. The temperature dependence of the QE
line width then follows
\begin{equation}
\rm \Gamma/2(T) = \Gamma_0/2 + c \cdot \Delta_{CF}^3 /(e^{\Delta_{CF}/T}-1)
\label{gleichung}
\end{equation}
where $\rm \Delta_{CF}$ is the energy of an excited crystal field
state and c is a factor which among others
considers the coupling of the CF ground state with the excited
CF state. In equation (\ref{gleichung}) only for $\rm T \gg \Delta_{CF}$ the
line width is proportional to the temperature.
We fit this function to the experimental data of many different samples and
obtained values around 200~K for $\rm \Delta_{CF}$. Since the energy of the
excited CF state should not differ much between our samples, we took a
mean value of $\rm \Delta_{CF}=200~K$ in the following. Unfortunately, the
crystal field scheme for Nd doped LSCO has yet not been
evaluated. However, $\rm \Delta_{CF}\approx 200~K$ coincides roughly
with the energy of the first excited state ($\rm \Delta_{CF}=173~K$) in
$\rm Nd_2CuO_4$\cite{Loong}.

A fit with equation (\ref{gleichung}) reveals $\rm c=1.54\cdot
10^{-7}~meV/K^3$ and $\rm c=1.93\cdot 10^{-7}~meV/K^3$ for
$\rm La_{1.25}Sr_{0.15}Nd_{0.6}CuO_4$ and $\rm La_{1.5}Sr_{0.2}Nd_{0.3}CuO_4$,
respectively. A residual line width of $\rm \Gamma_0/2=10 \mu eV$ has also been
taken into account but does not influence the other parameters
markedly\cite{blablabla}. The values for c are in the expected range\cite{Orbach}
and are in good agreement with the reported value for LSCO probed with
$\rm Er^{3+}$ spins\cite{Shimizu1}.\par
We now turn to a discussion of the relevance of the Orbach relaxation process in
other systems. A comparison of the QE line widths in $\rm La_{2-x-y}Sr_xNd_yCuO_4$
with the
reported values in $\rm Nd_2CuO_4$\cite{Casalta,Loewenhaupt2} and
$\rm NdBa_2Cu_3O_{7-\delta}$\cite{Droessler} shows that
the absolute values of $\rm \Gamma/2$ are of the same order of magnitude
in all compounds. For $\rm Nd_2CuO_4$
only a few data points of the QE line width exist. The agreement between
the data of Casalta et al. for a single crystal and Loewenhaupt et al. for a
powder sample is rather poor. A fit with the above function yields
$\rm \Gamma_0/2=0.24~meV$, $\rm c=1.42\cdot10^{-7}~meV/K^3$ and
$\rm \Gamma_0/2=0.17~meV$, $\rm c=8.5\cdot10^{-8}~meV/K^3$ for the data of
Loewenhaupt et al. and Casalta et al., respectively ($\rm \Delta_{CF}$ was
fixed at 175~K). Note that in both cases a large residual line width
is obtained from the fit. This broad Lorentzian line is probably masked by
the broad Gaussian line.
A large value of $\rm \Gamma_0/2$ was directly observed in $\rm
NdBa_2Cu_3O_{7-\delta}$\cite{Droessler}. A fit of the data on $\rm NdBa_2Cu_3O_6$
with eq.~(\ref{gleichung})
reveals $\rm c=3.35\cdot 10^{-8}~meV/K^3$ ($\rm \Gamma_0/2$ and $\rm \Delta_{CF}$
fixed at 0.235~meV and 410~K\cite{Allenspach1}, respectively). Two things are
worth mentioning. Firstly, the choice of 410~K for $\rm \Delta_{CF}$ has
the consequence that the increase of the line width is suppressed up to higher
temperatures. This is indeed observed in $\rm
NdBa_2Cu_3O_{7-\delta}$\cite{Droessler}. Secondly, c is roughly an order of
magnitude smaller than in $\rm La_{2-x-y}Sr_xNd_yCuO_4$. This is in agreement
with the observations of Shimizu et al. for an $\rm Er^{3+}$ spin probe in
LSCO and YBCO\cite{Shimizu1}. The broad residual line width in the concentrated
systems might be caused by a strongly enhanced Nd--Nd interaction.\par
A QE Lorentzian was also observed in $\rm Pb_2Sr_2TbCu_3O_8$\cite{Staub1} and
$\rm Pb_2Sr_2Tb_{0.5}Ca_{0.5}Cu_3O_8$\cite{Staub2}. In both compounds
Tb has a quasi--doublet ground state, i.e.  two singlets separated
by only a few $\rm \mu eV$. It was found that in $\rm
Pb_2Sr_2Tb_{0.5}Ca_{0.5}Cu_3O_8$ the temperature dependence of
$\rm \Gamma(T)$ obeys a power law
$\rm t^\nu$ with $\rm \nu=2.8 \; [t=(T-T_N)/T_N]$.
Conclusively the authors claimed that the rare--earth exchange
interaction might be the dominant process for the 4f--spin
relaxation. The same group obtained similar results for $\rm
Y_{0.9}Tb_{0.1}Ba_2Cu_3O_7$\cite{Staub3}. Since the concentration of Tb in
this compound is very low the above interpretation of a strong
rare--earth exchange interaction seems rather unlikely.
In contrast, a very recent reanalysis of the data showed that the
temperature dependence of the QE line width also follows the
two--phonon Orbach process\cite{Staub4}. Taking all these facts
into account it seems reasonable to assume that the 4f spin relaxation
in high $\rm T_c$ superconductors and related materials is caused by CF
transitions assisted by phonons and {\em not} by interaction with conduction
electrons. Moreover, the interpretation of the deviation from a linear
temperature dependence of $\rm \Gamma/2(T)$ as opening of a gap has to
be reexamined\cite{Boot,Mesot}.\par
To conclude this section we want to mention that a coupling between
these two elementary excitations was already reported in
Raman scattering studies on several high $\rm T_c$ superconductors and related
compounds\cite{Heyen1}. Furthermore, ESR data have been discussed in terms of
an Orbach process\cite{Shimizu1,Kan}. These data are consistent with our
interpretation of the temperature dependence of the QE line width
(see also our previous work~\cite{Roepke1}).

\subsection{Magnetic response due to Nd--Cu interaction}

In contrast to the above described behavior the 4f
magnetic response at low temperatures correlates with the electronic
properties of the $\rm CuO_2$--layers. Depending on the dopant concentration
we find different features of the magnetic response which we will now discuss in
detail.\\

{\bf x = 0}\\

\begin{figure}
\begin{center}
\includegraphics[clip, angle =270,width=8.5cm]{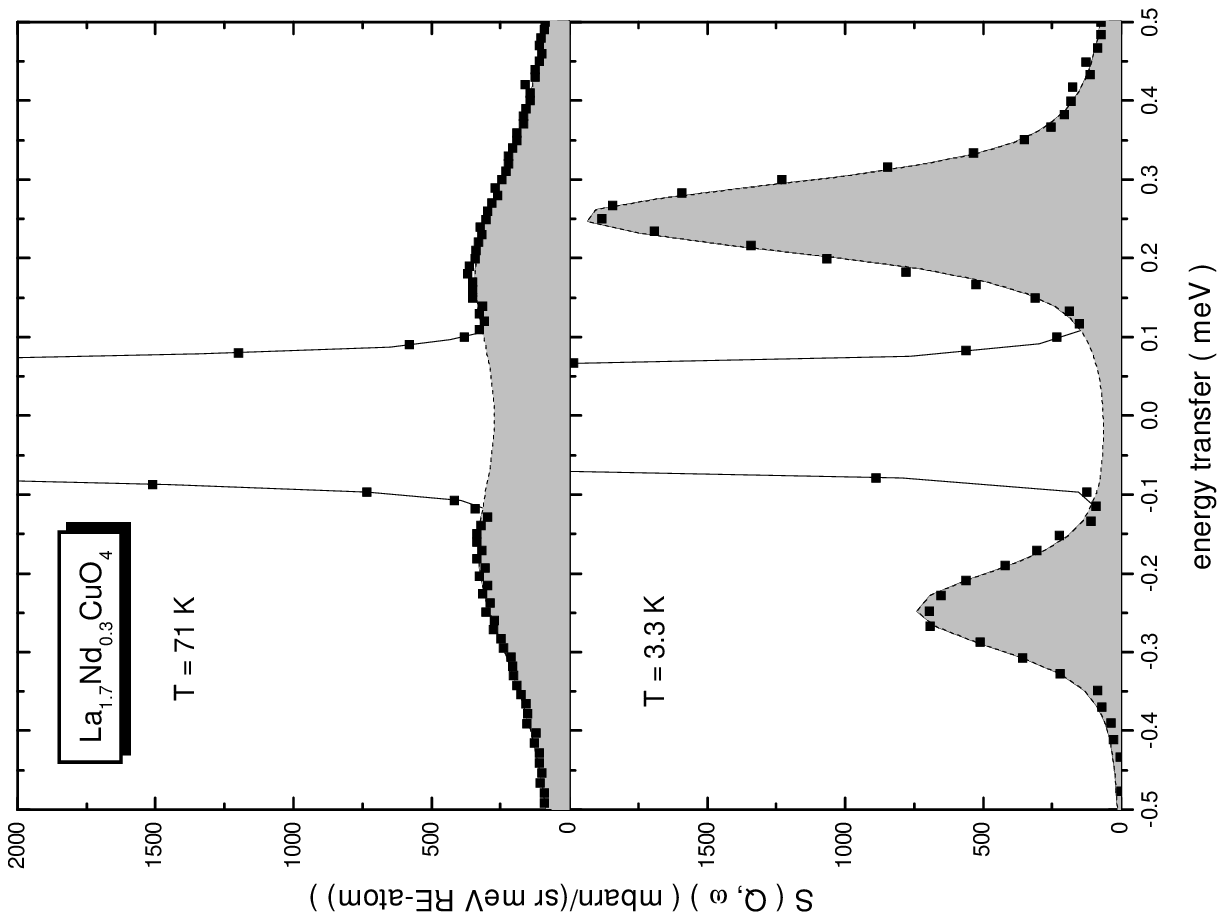}
\end{center}
\vspace*{-.15cm}
\begin{minipage}{8.5cm}
\caption{\it Background corrected spectrum of $La_{1.7}Nd_{0.3}CuO_4$ at T = 71~K
(MIBEMOL) and T = 3.3~K (NEAT). The full line is the best fit including the nuclear
incoherent elastic scattering. The magnetic contribution (INE Lorentzian) is given
by the shaded area.}
\end{minipage}
\label{prb003}
\end{figure}

In insulating $\rm La_{2-y}Nd_yCuO_4$ with y = 0.1, 0.3 the 4f magnetic
response changes from a QE Lorentzian to an inelastic (INE) excitation below
about 80~K\cite{Roepke} (see Fig. \ref{prb003}).
This INE excitation clearly indicates the splitting of the $\rm Nd^{3+}$ ground
state Kramers doublet due to the internal exchange field of ordered Cu moments
and shows the strong interaction between the Cu-- and Nd--subsystems.
Although the Cu ordering temperature is much higher, this excitation becomes first
detectable below 80~K since at higher temperature the line width is larger than
the observed energy splitting. We mention that this is in contrast to
neutron scattering results
on $\rm Nd_2CuO_4$ where the magnetic signal remains QE down to
5~K\cite{Loewenhaupt2} possibly due to Nd--Nd interactions.
For the sample with smaller Nd content, namely $\rm La_{1.9}Nd_{0.1}CuO_4$,
we find a similar behavior as for y = 0.3. In contrast to y = 0.3 in this
sample only a minority fraction of about 20 \% changes to
the Pccn phase\cite{Cramm}.

\begin{figure}
\begin{center}
\includegraphics[clip, angle =270,width=8.5cm]{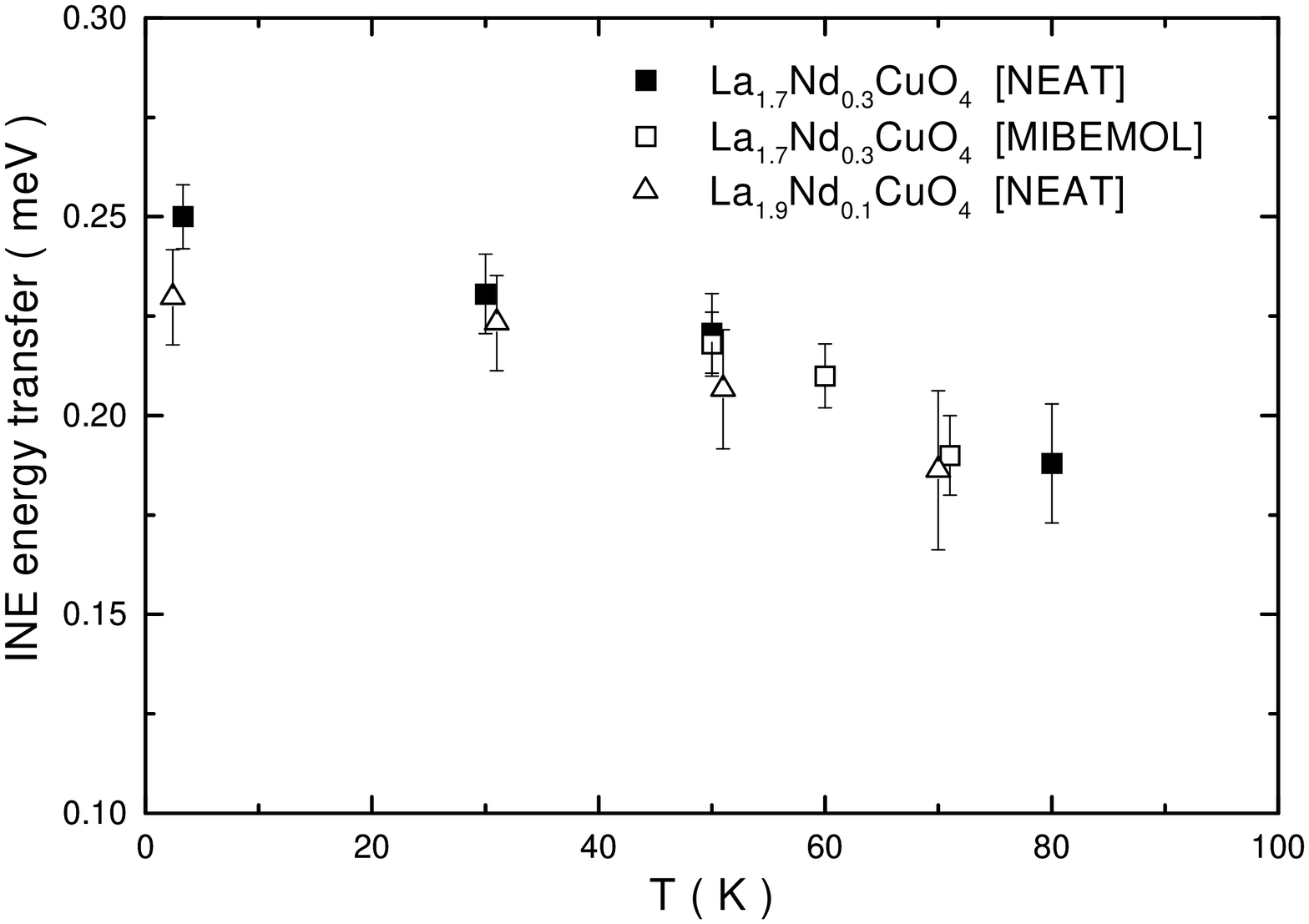}
\end{center}
\vspace*{-.15cm}
\begin{minipage}{8.5cm}
\caption{\it Temperature dependence of the energy excitation in
$La_{2-y}Nd_yCuO_4$ for y = 0.1 and y = 0.3.}
\end{minipage}
\label{prb004}
\end{figure}

The temperature dependence of the energy splitting
for both compounds is plotted in Fig. \ref{prb004}. An increase of the
splitting, i.e. of the internal exchange field at the Nd site, is clearly
visible. Our findings agree roughly with the data of Chou et al.\cite{Chou} who
measured the internal field at the La site in $\rm La_2CuO_4$ with
$\rm^{139}La$ NQR. According to their results the temperature dependence of
the internal field can be
described with a power law $\rm (1-T/T_N)^{\beta}$ ($\rm T_N=250~K$ and $\rm
\beta=0.41$). This means an increase of about 17\% from 80~K to 3~K and agrees
roughly with our results in $\rm La_{1.9}Nd_{0.1}CuO_4$. It is obvious that the
exchange field at the Nd site is influenced by both, the staggered magnetization
and the direction of the Cu spins. In $\rm La_{1.7}Nd_{0.3}CuO_4$
the structural transition from $\rm LTO \to Pccn$ is accompanied by a Cu spin
reorientation\cite{Crawford,Keimer}. Thus, the enhanced splitting in y = 0.3
compared to y = 0.1 for $\rm T \to 0~K$ might be due to the difference in the
direction of the Cu spins. Finally, we want to mention that the value of $\rm
\Delta E$ coincides with that derived from the Schottky anomaly found in
low temperature specific heat measurements\cite{Nguyen}.\\

{\bf x = 0.12}\\

Static ordering of charges and spins was first reported in $\rm
La_{1.48}Sr_{0.12}Nd_{0.4}CuO_4$\cite{Tranquada1}. Similar results are obtained
for x = 0.15\cite{Tranquada2} where magnetic order has also been observed with
$\rm \mu^+SR$--experiments\cite{Wagener1} and M\"ossbauer
experiments\cite{Breuer2}.

\begin{figure}
\begin{center}
\includegraphics[clip, angle =270,width=8.5cm]{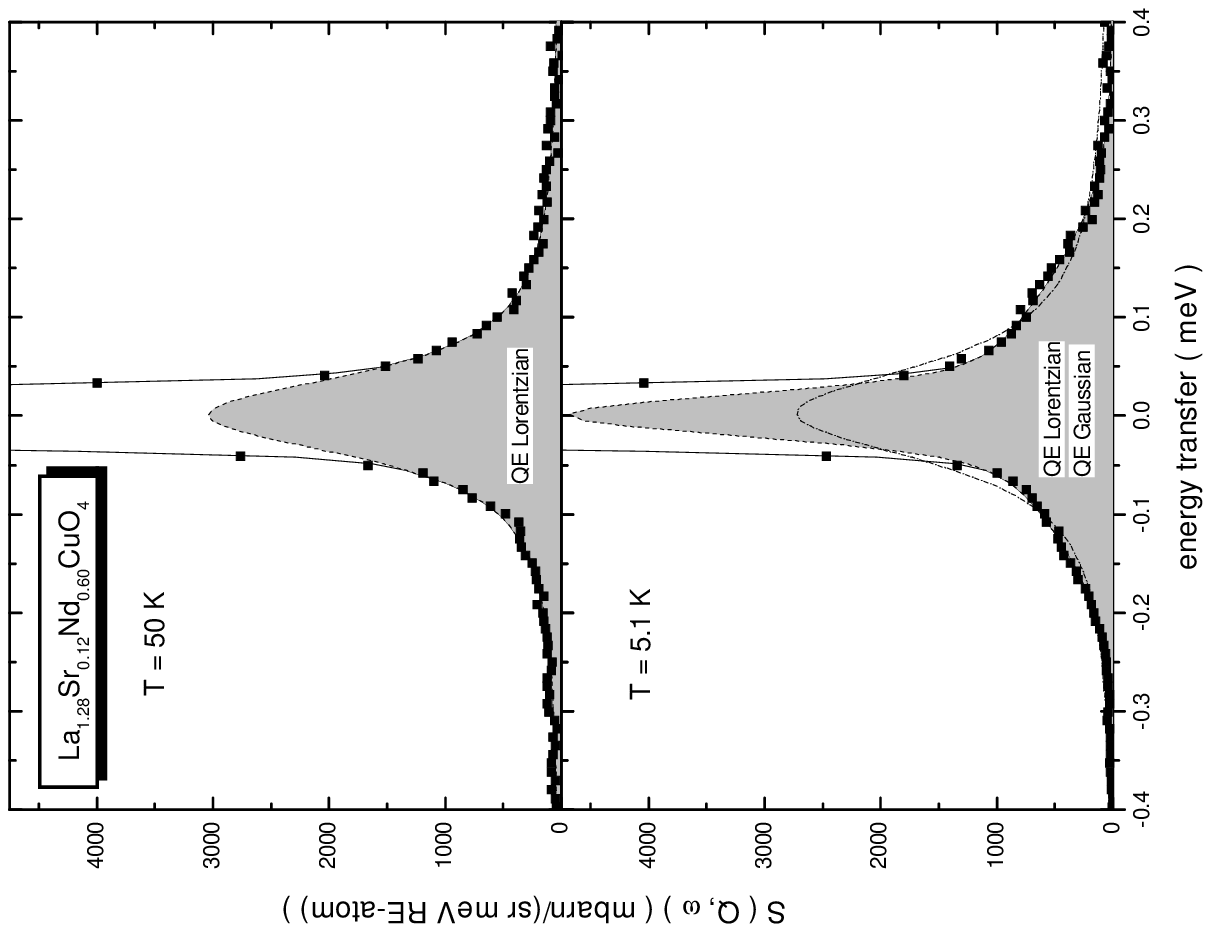}
\end{center}
\vspace*{-.15cm}
\begin{minipage}{8.5cm}
\caption{\it\hfill Background \hfill corrected \hfill spectrum \hfill of\protect\newline
$La_{1.28}Sr_{0.12}Nd_{0.6}CuO_4$
(MIBEMOL). The full line is the best fit including the
nuclear incoherent elastic scattering. The magnetic contribution is illustrated by the
shaded area. The dotted--dashed line shows the fit assuming a single Lorentzian.}
\end{minipage}
\label{prb004a}
\end{figure}

\begin{figure}
\begin{center}
\includegraphics[clip, angle =270,width=8.5cm]{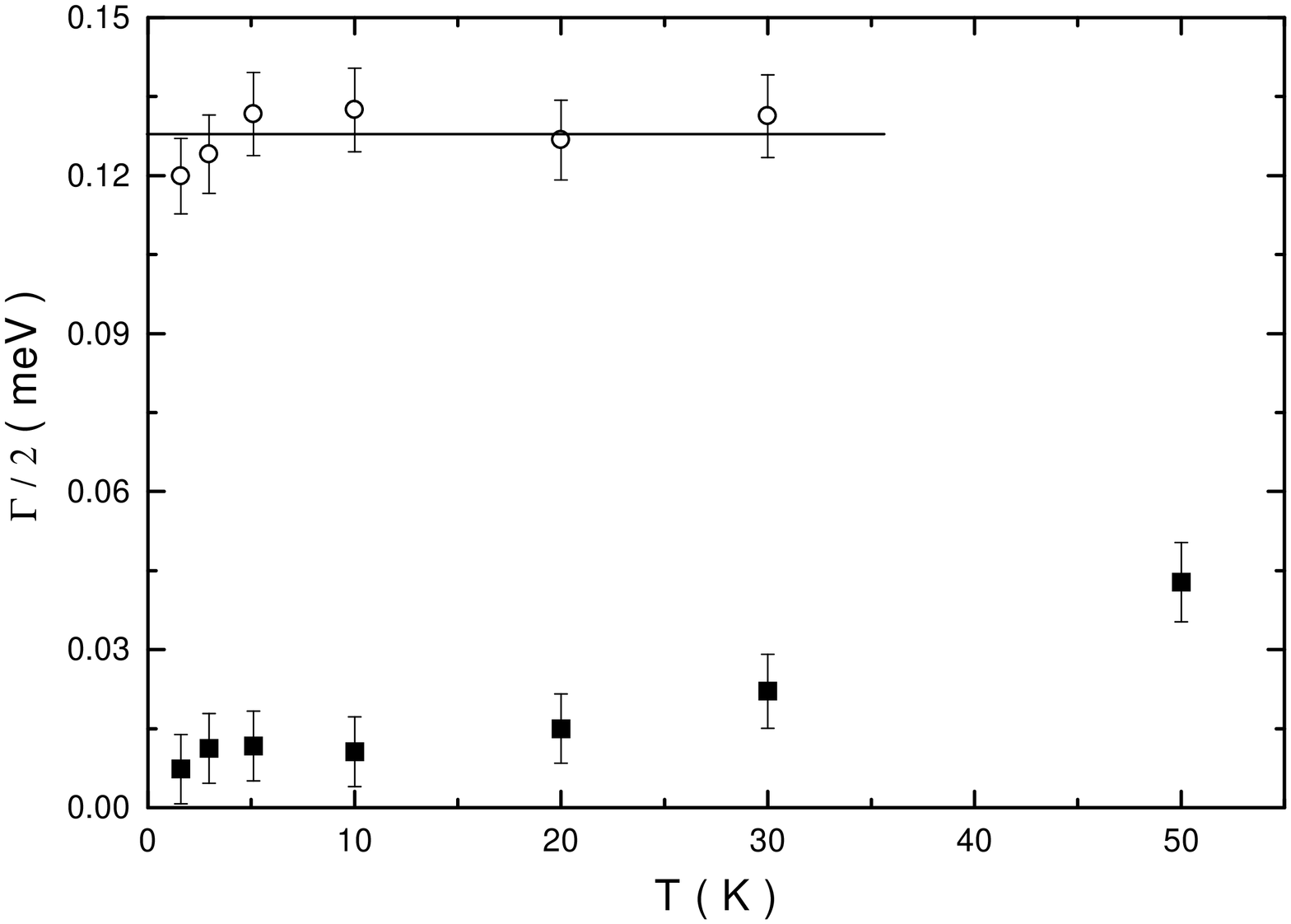}
\end{center}
\vspace*{-.15cm}
\begin{minipage}{8.5cm}
\caption{\it\hfill Quasielastic \hfill line \hfill widths \hfill (HWHM) \hfill
in\protect\newline $La_{1.28}Sr_{0.12}Nd_{0.6}CuO_4$. Open circles are
associated with the Gaussian line. Closed squares symbolize Lorentzian line
widths.}
\end{minipage}
\label{prb005}
\end{figure}

We performed measurements on
several compounds related to this composition and found similar properties at
low temperatures. As a representative sample
we chose $\rm La_{1.28}Sr_{0.12}Nd_{0.6}CuO_4$ to explain the general
features that are observed when the type of ordering changes from the well
known spin structure for x = 0 into an antiferromagnetic stripe pattern of
ordered spins and charges in the $\rm CuO_2$--layers. Above the
antiferromagnetic ordering temperature ($\rm T_N \approx 30~K$) a single QE
Lorentzian line is found as discussed in section A. When the temperature is
further lowered below $\rm T_N$ the line width does not become smaller as
expected from the two--phonon Orbach process. Instead, a
broad magnetic response of almost constant width centered around the elastic
peak is visible (see Fig. \ref{prb004a}). The analysis of our results shows
that in addition to the Lorentzian a Gaussian line is necessary to accurately
describe the
data. The width of this Gaussian is almost temperature independent
($\rm \Gamma_{Gaussian}/2 \approx 0.13~meV$, see Fig. \ref{prb005}).

The observation of a broad QE Gaussian line instead of a well resolved INE
excitation as for x = 0 infers a distribution of different energy splittings
on different Nd sites. This is a direct hint on spatial inhomogeneities in
the $\rm CuO_2$--planes and is probably caused by the
formation of stripes. In this picture the Kramers doublet of a $\rm Nd^{3+}$
ion, which is located 'near' a charge stripe and thus is not split up,
contributes to the Lorentzian signal which is still observable (at least the
splitting must be small compared to the width of this Lorentzian). The
comparison of the width of the Gaussian line in $\rm x= 0.12$ with the energy
excitation of the inelastic line (x = 0) reveals a reduced (average) splitting,
which is related to a reduced zero temperature staggered magnetization in the
$\rm CuO_2$--planes. This finding coincides with the $\rm
\mu^+SR$--experiments\cite{Wagener1} on $\rm La_{1.85-y}Sr_{0.15}Nd_yCuO_4$
where a decrease of the muon spin rotation frequency compared to $\rm
La_{1.7}Nd_{0.3}CuO_4$\cite{Wagener2} was found. This was interpreted as a
decrease of the average magnetic field at the muon site in Sr doped
compounds.\par
\begin{figure}
\begin{center}
\includegraphics[clip, angle =270,width=8.5cm]{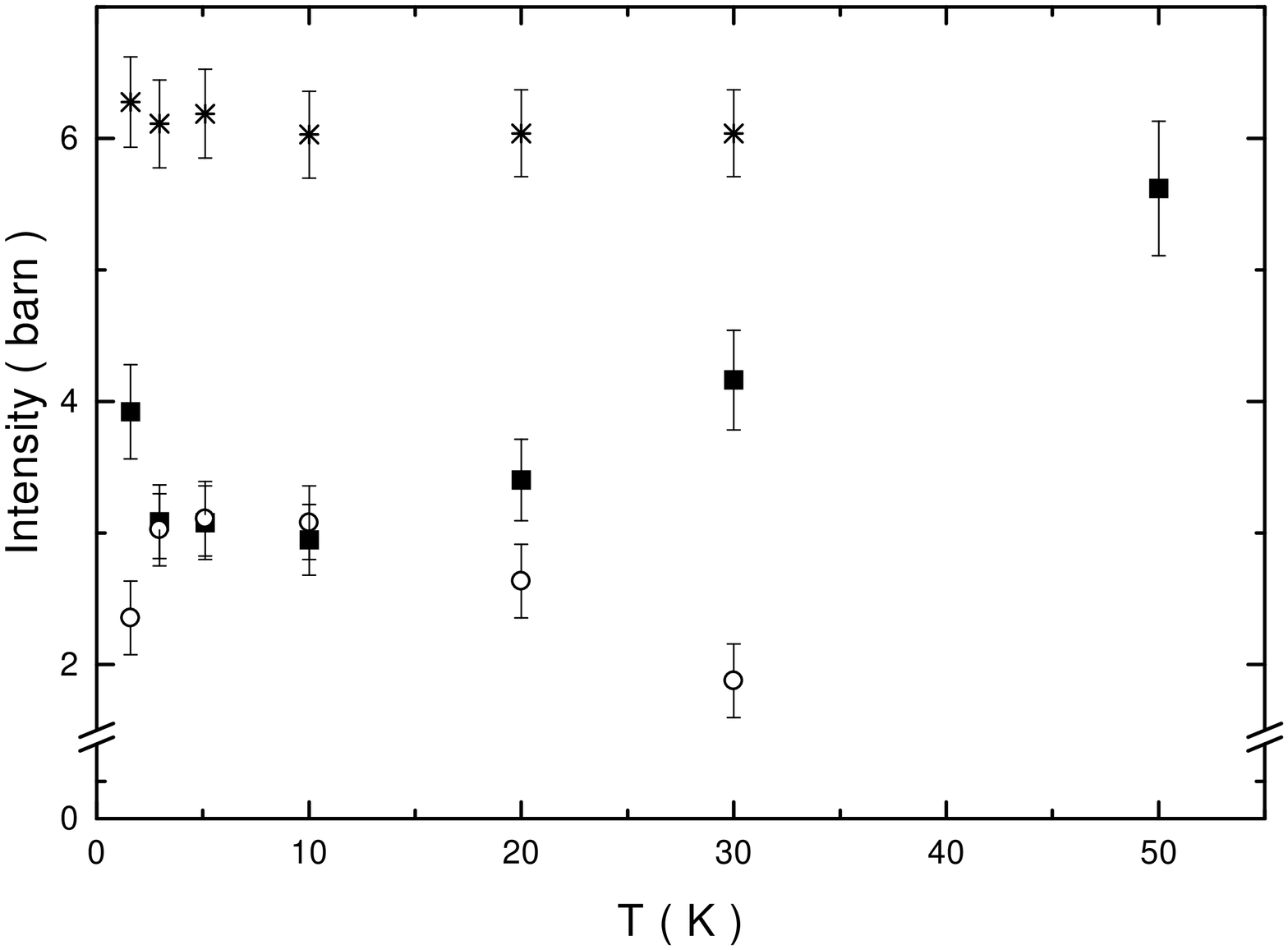}
\end{center}
\vspace*{-.15cm}
\begin{minipage}{8.5cm}
\caption{\it Intensity of Gaussian (open circles) and Lorentzian (closed squares)
line in $La_{1.28}Sr_{0.12}Nd_{0.6}CuO_4$. The asterisks are the overall
magnetic intensity.} \label{prb006}
\end{minipage}
\end{figure}

In Fig. \ref{prb006} the intensities of the Lorentzian and
Gaussian signals are plotted versus temperature. Above 50~K the intensity of the
single Lorentzian line raises with decreasing temperature due to an
increase in the thermal occupation of the ground state (not shown). Below this
temperature firstly the intensity of the Lorentzian decreases linearly
for $\rm T\,\gsim \, 10~K$ whereas
the intensity of the Gaussian raises. This can be interpreted as a decrease
in the number of paramagnetic Nd ions (i.e. with a splitting smaller than the
Lorentzian line width). This is expected since the
Lorentzian line width decreases with decreasing temperature and therefore the
number of Nd ions for which the above condition holds reduces. Secondly,
between 10~K and 3~K, a plateau--like level is reached. For $\rm T\,\lsim \,
10~K$ the
Lorentzian line width is well below the resolution limit and so small that we
cannot distinguish between a Lorentzian line and the elastic line.
Finally,
below 3~K a strong increase of the Lorentzian intensity (which we cannot
distinguish from an increase of the elastic intensity) is observed. At a similar
temperature a pronounced increase of the magnetic intensity due to ordering of
the Nd moments was reported in $\rm
La_{1.48}Sr_{0.12}Nd_{0.4}CuO_4$\cite{Tranquada1}.

{\bf Sr dependence of the magnetic response}\\

Besides the above mentioned two Sr--concentrations x = 0 (with y = 0.1, 0.3)
and x = 0.12 (with y = 0.6) experiments with varying Sr compositions were
carried out on $\rm La_{1.7-x}Sr_xNd_{0.3}CuO_4$ with x = 0.12, 0.15, 0.2 and
$\rm La_{1.4-x}Sr_xNd_{0.6}CuO_4$ with x = 0.1, 0.15, 0.18, 0.2.
For clarity the data of the series with y = 0.3 are
not shown since the observations in these compounds are similar to the findings
in the related samples with y = 0.6.\par
\begin{figure}
\begin{center}
\includegraphics[clip, angle =270,width=8.5cm]{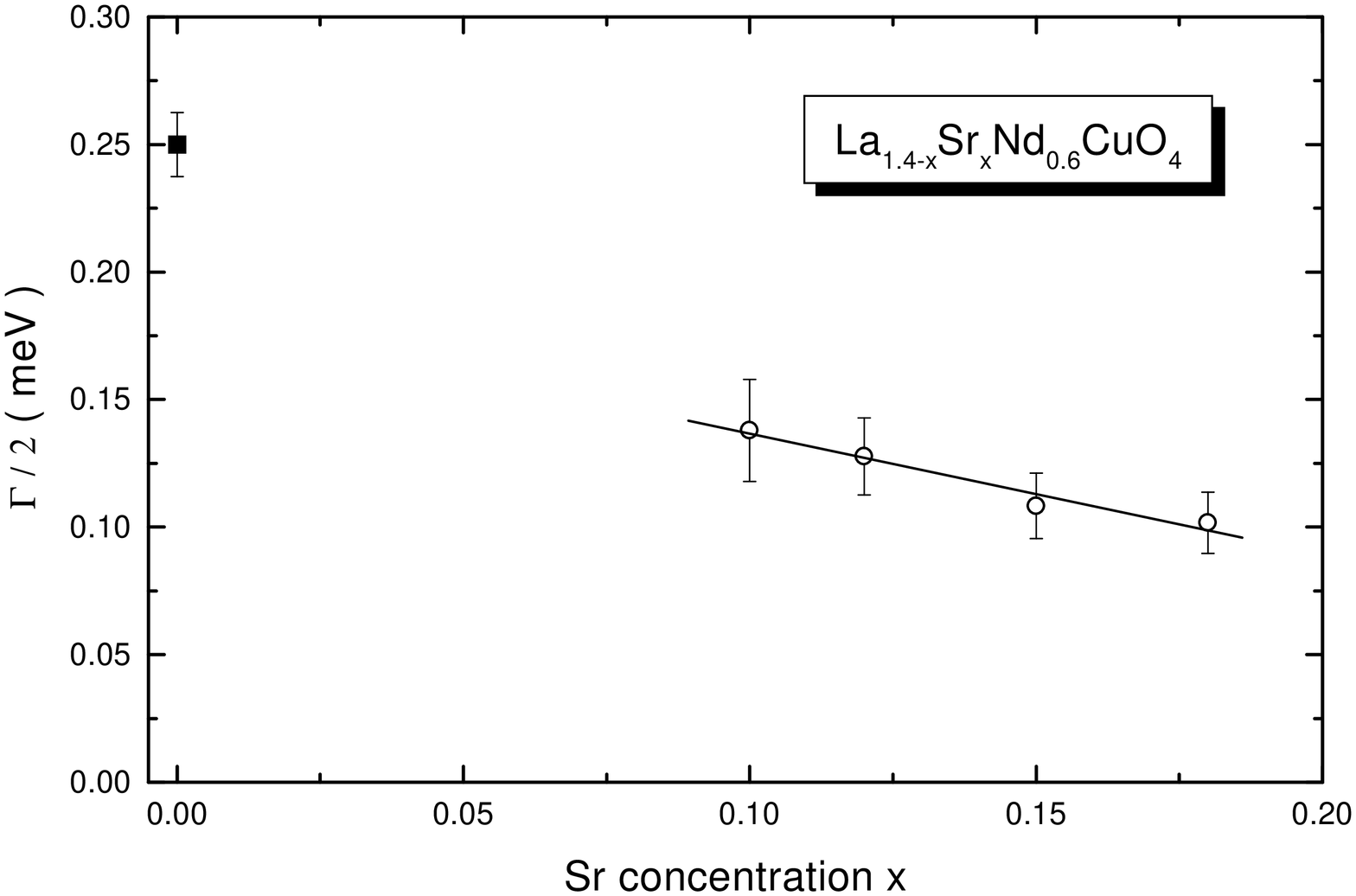}
\end{center}
\vspace*{-.15cm}
\begin{minipage}{8.5cm}
\caption{\it\hfill Gaussian \hfill line \hfill widths \hfill (open \hfill circles)
\hfill vs.\protect\newline
Sr--concentration. The solid line is a 'guide to the eye'. For comparison
the INE energy splitting of $La_{1.7}Nd_{0.3}CuO_4$ at 3.3~K is also plotted
(closed square).}
\end{minipage}
\label{prb007}
\end{figure}

To study the Sr dependence of the magnetic signal at low temperatures in detail
we extended our
measurements to samples with a Nd content of y = 0.6 with $\rm 0.1 \leq x \leq 0.2$.
In contrast to
$\rm La_{1.7-x}Sr_xNd_{0.3}CuO_4$ all of these samples lie in the region of the
phase diagram where
superconductivity is strongly suppressed\cite{Buechner} and thus a broad magnetic
response at low temperatures is expected. Such a response was found in all
samples except for $\rm La_{1.2}Sr_{0.2}Nd_{0.6}CuO_4$. The line widths of the
QE Gaussians are plotted in Fig. \ref{prb007} (these data points are fits under
the assumption that $\rm \Gamma_{Gaussian}/2$ is temperature independent for each
compound).
The decrease of the line width with increasing Sr content is related to a
reduction of the average staggered magnetization in the $\rm CuO_2$--planes.
This observation is consistent with the findings in $\rm
La_{1.6-x}Sr_xNd_{0.4}CuO_4$ with x = 0.12, 0.15 and 0.2\cite{Tranquada2}.

Unfortunately the spectra of $\rm La_{1.3}Sr_{0.1}Nd_{0.6}CuO_4$ showed
strongly enhanced background probably due the diffusion of air into the neutron
flight path. Nevertheless, the analysis showed that a QE
Gaussian is not consistent with the data in the whole temperature range 1.8~K --
30~K although it is accurate enough at certain temperatures. A better agreement
is obtained when we use an INE Gaussian line. This might reflect a mixture of both
types of signals which we observe in $\rm 0 < x = 0.1 < 0.12,\,0.15,\,0.18$,
i.e. an INE excitation and QE Gaussian, respectively. Furthermore we note that
this compound is closest to the Pccn phase.\par
\begin{figure}
\begin{center}
\includegraphics[clip, angle =270,width=8.5cm]{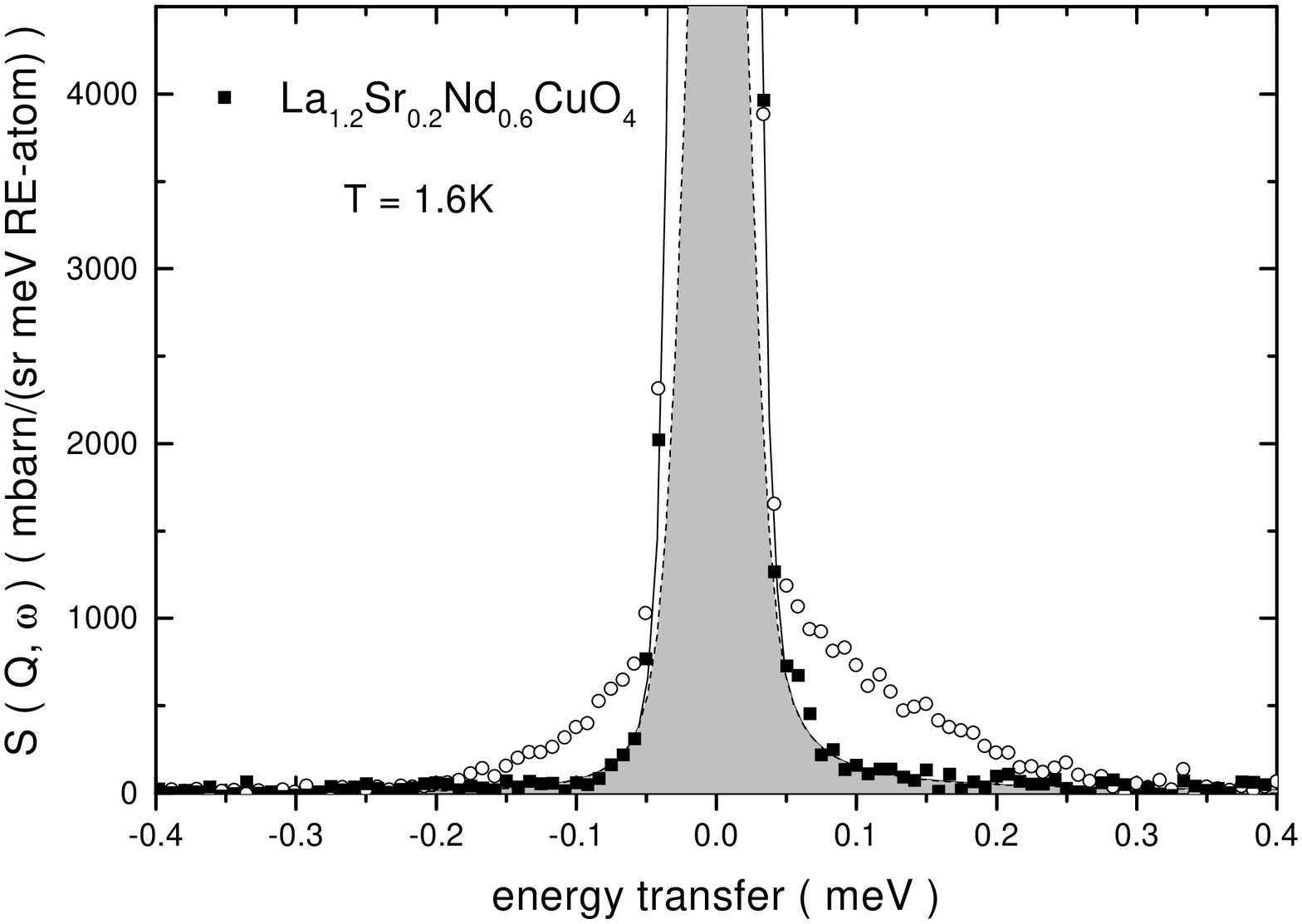}
\end{center}
\vspace*{-.15cm}
\begin{minipage}{8.5cm}
\caption{\it\hfill Background \hfill corrected \hfill spectrum \hfill of\protect\newline
$La_{1.2}Sr_{0.2}Nd_{0.6}CuO_4$ at 1.6~K. The magnetic contribution (QE
Lorentzian) is given
by the shaded area. The open circles represent the magnetic
response which is found in $La_{1.28}Sr_{0.12}Nd_{0.6}CuO_4$ at the same
temperature. We compare these samples since both were measured on the same
spectrometer (MIBEMOL). Due to the Bose factor the spectral weight is shifted
asymmetrically to the neutron energy loss side.} \label{prb008}
\end{minipage}
\end{figure}

We could not detect any QE broadening in $\rm La_{1.8-y}Sr_{0.2}Nd_yCuO_4$ with
y = 0.3 and 0.6 at lowest temperature (see Fig. \ref{prb008}). For y = 0.3 this
behavior is expected since this compound is a bulk superconductor below $\rm
T_c \approx 25~K$ and hence no magnetic order in the $\rm CuO_2$--planes is
expected. Indeed, in a superconductor with even higher Nd--concentration
($\rm La_{1.4}Sr_{0.2}Nd_{0.4}CuO_4$) Nachumi et
al. could not find any hints for magnetic order in a recent $\rm
\mu SR$--experiment\cite{Nachumi}.
In contrast to this, the absence of a QE broadening is surprising in the compound
with y = 0.6 because superconductivity is strongly suppressed and hence a broad
magnetic response is expected.
Our sample has also been studied in a  recent $\rm \mu SR^+$--experiment
which shows magnetic order below about 15~K\cite{Klauss}.
This finding seems to contradict our neutron scattering results.
There might be two reasons why a broadening is not observable in the
present neutron scattering experiment\,: (i) the splitting of the Nd Kramers
ground state is too small or (ii) the intensity of the QE Gaussian is too low.
From the energy resolution chosen in our experiment we can conclude that a
possible average splitting of the Nd ground state must be below $\rm \approx
20 \mu eV$ if case (i) were true. This contradicts the value which is obtained
by extrapolating the Gaussian widths of Fig. \ref{prb007} to x = 0.2. Indeed,
scenario (ii) is more likely since we observe a drastic drop of the intensity
of the QE Gaussian in y = 0.18 compared to y = 0.15 at lowest temperature.
At present we have no interpretation for this drop of the intensity. Further
combined neutron and $\rm \mu SR$ studies are necessary in order to investigate
and understand this pronounced concentration dependence of the intensity.

\section{Summary}

To summarize, we have presented inelastic magnetic neutron scattering
experiments on Nd doped $\rm La_{2-x}Sr_xCuO_4$. In all samples at higher
temperatures a quasielastic line of Lorentzian shape is observed with a line
width which decreases with decreasing temperature. The temperature dependence
of this width, i.e. the relaxation of the Nd 4f--moments is dominated by the
Orbach relaxation process via the coupling of phonons and CF excitations and
does not depend on the charge carrier concentration in the $\rm CuO_2$--planes.
The low temperature behavior
of the magnetic response clearly correlates with the electronic properties of
the $\rm CuO_2$--layers. In the undoped samples (x = 0) below about
80~K an inelastic excitation occurs which shows the splitting of the $\rm
Nd^{3+}$ Kramers doublet ground state due to the Cu exchange field at the Nd
site. In $\rm La_{1.7-x}Sr_xNd_{0.3}CuO_4$ with x = 0.12, 0.15 and $\rm
La_{1.4-x}Sr_xNd_{0.6}CuO_4$ with x = 0.1, 0.12, 0.15, 0.18 superconductivity
is strongly suppressed. In all these compounds we observe an additional
quasielastic Gaussian below about 30~K. The width of this Gaussian is almost
temperature independent and decreases with increasing Sr concentration. The
observation of a Gaussian line infers a distribution of various Cu exchange
fields on different Nd sites and is interpreted in terms of the
stripe model. In $\rm La_{1.8-y}Sr_{0.2}Nd_yCuO_4$ (y = 0.3, 0.6) no indication
for a Nd--Cu interaction has been found, i.e. a single quasielastic Lorentzian
is observed.

\section{Acknowledgment}

We gratefully acknowledge useful discussions with V. Kataev and M.M.
Abd--Elmeguid.
Our work was supported by the 'BMBF' under contract number 03-HO4KOE.

\end{multicols}

\begin{thebibliography}{10}
%
\bibitem{Incommens} S.--W. Cheong, G. Aeppli, T.E. Mason, H.A. Mook, S.M. Hayden,
P.C. Canfield, Z. Fisk, K.N. Klausen and J.L. Martinez, {\sl Phys. Rev. Lett.}
{\bf 67}, 1791 (1991)\\
M. Matsuda, K. Yamada, Y. Endoh, T.R. Thurston, G. Shirane, R.J. Birgenau, M.A.
Kastner, I. Tanaka and H. Kojima, {\sl Phys. Rev. B} {\bf 49}, 6958 (1994)\\
K. Yamada, S. Wakimoto, C.H. Lee, M.A. Kastner, S. Hosaya, M. Greven, Y. Endoh
and R.J. Birgenau, {\sl Phys. Rev. Lett.} {\bf 75}, 1626 (1995)\\
K. Yamada, C.H. Lee, K. Kurahashi, J. Wada, S. Wakimoto, S.
Ueki, H. Kimura and Y. Endoh, {\sl Phys. Rev. B} {\bf 57}, 6165 (1998)\\
S. Petit, A.H. Moudden, B. Hennion, A. Vietkin and A.
Revcolevschi, {\sl Eur. Phys. J. B} {\bf 3}, 163 (1998)
\bibitem{Tranquada1} J.M. Tranquada, B.J. Sternlieb, J.D. Axe, Y. Nakamura and
S. Uchida, {\sl Nature} {\bf 375}, 561 (1995)\\
J.M. Tranquada, J.D. Axe, N. Ichikawa, Y. Nakamura, S. Uchida and B. Nachumi
{\sl Phys. Rev. B} {\bf 54}, 7489 (1996)
\bibitem{Crawford} M.K. Crawford, R.L. Harlow, E.M. McCarron, W.E. Farneth and
N. Herron, {\sl Phys. Rev. B} {\bf 47}, 11623 (1993)
\bibitem{Buechner}  B. B\"uchner, M. Breuer, A. Freimuth and A.P. Kampf
{\sl Phys. Rev. Lett.} {\bf 73}, 1841 (1994)
\bibitem{Dai} P. Dai, H.A. Mook and F. Dogan, {\sl Phys. Rev. Lett.} {\bf 80},
1738 (1998)
\bibitem{Henggeler1} W. Henggeler, T. Chattopadhyay, P. Thalmeier, P.
Vorderwisch and A. Furrer, {\sl Europhys. Lett.} {\bf 34}, 537 (1996)
\bibitem{Ivanov} A.S. Ivanov, P. Bourges, D. Petitgrand and J. Rossat--Mignod
{\sl Physica B} {\bf 213 \& 214}, 60 (1995)
\bibitem{Casalta} H. Casalta, P. Bourges, M. d'Astuto, D. Petitgrand and
A. Ivanov, {\sl Phys. Rev. B} {\bf 57}, 471 (1998)
\bibitem{Loewenhaupt1} M. Loewenhaupt, A. Metz, N.M. Pyka, D. McK. Paul, J.
Martin, V.H.M. Duijn, J.J.M. Franse, H. Mutka and W. Schmidt, {\sl Ann. Physik}
{\bf 5}, 197 (1996)
\bibitem{Henggeler2} W. Henggeler, T. Chattopadhyay, B. Roessli, P.
Vorderwisch, P. Thalmeier, D.I. Zhigunov, S.N. Barilo and A. Furrer, {\sl
Phys. Rev. B} {\bf 55}, 1269 (1997)
\bibitem{Maiser} T. Brugger, T. Schreiner, G. Roth, P. Adelmann and G. Czjzek,
{\sl Phys. Rev. Lett.} {\bf 71}, 2481 (1993)
\bibitem{Henggeler3} W. Henggeler, B. Roessli, A. Furrer, P.
Vorderwisch, T. Chatterji, {\sl Phys. Rev. Lett.} {\bf 80}, 1300 (1998)
\bibitem{Markert} J.T. Markert, E.A. Early, T. Bjornholm, S. Ghamaty, B.W. Lee,
J.J. Neumeier, R.D. Price, C.L. Seaman and M.B. Maple, {\sl Physica C} {\bf
158}, 187 (1989)
\bibitem{Jandl1} S. Jandl, M. Iliev, C. Thomsen, T. Ruf and M. Cardona
{\sl Solid State Comm.} {\bf 87}, 609 (1993)
\bibitem{Dufour} P. Dufour, S. Jandl, C. Thomsen, M. Cardona, B.M. Wanklyn
and C. Changkang, {\sl Phys. Rev. B} {\bf 51}, 1053 (1995)
\bibitem{Jandl2} S. Jandl, P. Dufour, T. Strach, T. Ruf, M. Cardona, V.
Nekvasil, C. Chen, B.M. Wanklyn and S. Pinol, {\sl Phys. Rev. B} {\bf 53},
8632 (1996)
\bibitem{Loewenhaupt2} M. Loewenhaupt, P. Fabi, S. Horn, P. v. Aken and
A. Severing, {\sl J. Magn. Magn. Mater.} {\bf 140--144}, 1293 (1995)
\bibitem{Droessler} H. Dr\"ossler, H.D. Jostarndt, J. Harnischmacher,
J. Kalenborn, U. Walter, A. Severing, W. Schlabitz and E. Holland--Moritz
{\sl Z. Phys. B} {\bf 100}, 1 (1996)

\bibitem{Lechner} R. E. Lechner, {\sl Neutron News} 7, {\bf No. 4} 9-11 (1996)
\bibitem{Report} H. D. Jostarndt and U. Walter, {\sl ILL report to experiment
4--05--230} (1989)
\bibitem{Breuer} M. Breuer, B. B\"uchner, R. M\"uller, M. Cramm, O. Maldonado,
A. Freimuth, B. Roden, R. Borowski, B. Heymer and D. Wohlleben, {\sl Physica C}
 {\bf 208}, 217 (1993)
\bibitem{Holland} E. Holland--Moritz and G. Lander, {\sl Handbook of the
Chemistry and Physics of Rare Earths} {\bf 19}, 1 (1994)

\bibitem{Walter} U. Walter, S. Fahy, A. Zettl, S.G. Louie, M.L. Cohen, P.
Tejedor and A.M. Stacy, {\sl Phys. Rev. B} {\bf 36}, 8899 (1987)
\bibitem{Allenspach} P. Allenspach, A. Furrer and F. Hulliger, {\sl Phys.
Rev. B} {\bf 39}, 2226 (1989)
\bibitem{Allenspach2} P. Allenspach, A. Furrer, P. Bruesch and P. Unternaehrer,
{\sl Physica B} {\bf 156 \& 157}, 864 (1989)
\bibitem{Becker} K.W. Becker, P. Fulde and J. Keller,
{\sl Z. Phys. B} {\bf 28}, 9 (1977)
\bibitem{Fulde} P. Fulde and M. Loewenhaupt, {\sl Adv. Phys.} {\bf 34}, 589
(1986)
\bibitem{Staub2} U. Staub, L. Soderholm, S. Skanthakumar, S. Rosenkranz,
C. Ritter and W. Kagunya, {\sl Europhys. Lett.} {\bf 34}, 447 (1996)
\bibitem{Orbach} A. Abragam, B. Bleaney, Electron Paramagnetic Resonance of
Transition Ions, Clarendon, Oxford, ch. 10 (1970)
\bibitem{Loong} C.--K. Loong and L. Soderholm, {\sl Phys. Rev. B} {\bf 48},
14001 (1993)
\bibitem{blablabla} We do not know if there is really a difference in the relaxation
rates (and thereby in the fit parameters) of our samples. The
data points at high temperatures $\rm T>200~K$ have a large uncertainty due to
the fact that the signal to noise ratio is rather bad.
Furthermore, the Nd content in $\rm La_{1.7-x}Sr_xNd_{0.3}CuO_4$ is low
resulting in a poor statistics. Hence, the coincidence of the line widths of
$\rm La_{1.7}Nd_{0.3}CuO_4$ and $\rm La_{1.55}Sr_{0.15}Nd_{0.3}CuO_4$ at
240~K might be accidental. The most accurate results we obtained for $\rm
La_{1.25}Sr_{0.15}Nd_{0.6}CuO_4$.

\bibitem{Shimizu1} H. Shimizu, K. Fujiwara and K. Hatada, {\sl Physica C}
{\bf 288}, 190 (1997)
\bibitem{Allenspach1} P. Allenspach, J. Mesot, U. Staub, M. Guillaume, A.
Furrer, S.--I. Yoo, M.J. Kramer, R.W. McCallum, H. Maletta, H. Blank, H. Mutka,
R. Osborn, M. Arai, Z. Bowden and A.D. Taylor, {\sl Z. Phys. B} {\bf 95},
301 (1994)
\bibitem{Staub1} U. Staub, L. Soderholm, S. Skanthakumar and M.R. Antonio
{\sl Phys. Rev. B} {\bf 52}, 9736 (1995)
\bibitem{Staub3} U. Staub, F. Fauth, M. Gutmann and W. Kagunya, {\sl Physica B}
 {\bf 234--236}, 841 (1997)
\bibitem{Staub4} U. Staub, {\sl private communication}.
\bibitem{Boot} A.T. Boothroyd, A. Mukherjee and A.P. Murani, {\sl Phys. Rev. Lett.}
{\bf 77}, 1600 (1996)
\bibitem{Mesot} J. Mesot, G. B\"ottger, P. Berastegui, H. Mutka and A. Furrer,
{\sl Physica C} {\bf 282 -- 287}, 1377 (1997)
\bibitem{Heyen1} E. T. Heyen, R. Wegerer and M. Cardona, {\sl Phys. Rev. Lett.}
{\bf 67}, 144 (1991)\\
E. T. Heyen, R. Wegerer, E. Sch\"onherr and M. Cardona, {\sl
Phys. Rev. B} {\bf 44}, 10195 (1991)\\
J. A. Sanjurjo, C. Rettori, S. Oseroff and Z. Fisk, {\sl
Phys. Rev. B} {\bf 49}, 4391 (1994)\\
R. Wegerer, C. Thomsen, T. Ruf, E. Sch\"onherr, M. Cardona,
M. Reedyk, J.S. Xue, J.E. Greedan and A. Furrer, {\sl Phys. Rev. B} {\bf 48},
6413 (1993)
\bibitem{Kan} L. Kan, S. Elschner and B. Elschner, {\sl Solid State Comm.}
{\bf 79}, 61 (1991)
\bibitem{Roepke1} M. Roepke, E. Holland--Moritz, B. B\"uchner, R. Borowski, R. Kahn,
R.E. Lechner, S. Longeville and J. Fitter, {\sl J. of
Superconductivity} -- to be published

\bibitem{Roepke} M. Roepke, E. Holland--Moritz, G. Coddens, J. Fitter and
R.E. Lechner, {\sl Physica B} {\bf 234--236}, 723 (1997)\\
M. Roepke, E. Holland--Moritz, R. M\"uller, G. Coddens, R.E. Lechner,
S. Longeville and J. Fitter, {\sl J. Phys. Chem. Solids} {\bf 59}, 2233 (1998)
%\bibitem{Hien} N.T. Hien, V.H.M. Duijn, J.H.P. Colpa, J.J.M. Franse, and A.A.
%Menovsky, {\sl Phys. Rev. B} {\bf 57}, 5906 (1998)
\bibitem{Cramm} M. Cramm, {\sl private communication}
\bibitem{Chou} F.C. Chou, F. Borsa, J.H. Cho, D.C. Johnston, A. Lascialfari, D.
R. Torgeson and J. Ziolo, {\sl Phys. Rev. Lett.} {\bf 71}, 2323 (1993)
\bibitem{Keimer} B. Keimer, R.J. Birgenau, A. Cassanho, Y. Endoh, M. Greven,
M.A. Kastner and G. Shirane, {\sl Z. Phys. B} {\bf 91}, 373 (1993)
\bibitem{Nguyen} B. B\"uchner, {\sl private communication}
\bibitem{Tranquada2} J.M. Tranquada, J.D. Axe, N. Ichikawa, A.R. Moodenbaugh, Y.
Nakamura and S. Uchida, {\sl Phys. Rev. Lett.} {\bf 78}, 338 (1997)
\bibitem{Wagener1} W. Wagener, H.H. Klauss, M. Hillberg, M.A.C. de Melo, M.
Birke and F.J. Litterst, {\sl Phys. Rev. B} {\bf 55}, 14761 (1997)
\bibitem{Breuer2} M. Breuer, B. B\"uchner, H. Micklitz, E. Baggio--Saitovitch,
I. Souza Azevedo, R. Scorzelli, M.M. Abd--Elmeguid, {\sl Z. Phys. B} {\bf 92},
331 (1993)
\bibitem{Wagener2} W. Wagener, H.H. Klauss, M. Hillberg, M.A.C. de Melo, M.
Birke, F.J. Litterst, E. Schrier, B. B\"uchner and H. Micklitz, {\sl Hyperfine
Interact.} {\bf 105}, 107 (1997)

\bibitem{Nachumi} B. Nachumi, Y. Fudamoto, A. Keren, K.M. Kojima, M. Larkin,
G.M. Luke, J. Merrin, O. Tchernyshyov, Y.J. Uemura, N. Ichikawa, M. Goto, H.
Takagi, S. Uchida, M.K. Crawford, E.M. McCarron, D.E. McLaughlin and R.H.
Heffner, {\sl Phys. Rev. B} {\bf 58}, 8760 (1998)
\bibitem{Klauss} H. H. Klauss, {\sl private communication}
%
\end{thebibliography}
\end{document}